\documentclass[aps,prl,twocolumn,groupedaddress]{revtex4-1}

\usepackage{graphicx}
\usepackage{amssymb}
\usepackage{amsmath}

\newcommand{\rs}{\rm \scriptscriptstyle}

\begin{document}

\title {Collective many-body interaction in Rydberg dressed atoms}

\author{Jens Honer$^{1}$}
\author{Hendrik Weimer$^{1}$}
\author{Tilman Pfau$^{2}$}
\author{Hans Peter B\"uchler$^{1}$}
\affiliation{$^{1}$Institute for Theoretical Physics III, University of Stuttgart, Germany}
\affiliation{$^{2}$5. Physikalische Institut, University of Stuttgart, Germany}

\date{\today}

\begin{abstract}
We present a method to control the shape and character of the interaction potential between cold atomic gases by weakly dressing 
the atomic ground state  with a Rydberg level.  For increasing particle densities, a crossover takes place from a two-particle interaction
into a collective many-body interaction, where  the dipole-dipole/van der Waals Blockade phenomenon between the Rydberg levels plays a dominant role.
%
We study the influence of these collective interaction potential on a Bose-Einstein condensate, and present the optimal  parameters for its experimental detection.


\end{abstract}


\maketitle


The remarkable success of ultra-cold atomic  gases in exploring quantum
phenomena from the weakly interacting regime 
to strongly correlated many-body physics is founded in the microscopic
understanding of the two-body interaction potential, and the possibility to
control it by external fields.  
Several tools are nowadays available to
control the strength and shape of the interaction potential with the most
prominent examples beeing magnetic and optical Feshbach resonances 
\cite{tiesinga93,andrews98,courteille98,fedichev96}, 
atomic gases with large
magnetic dipole moments \cite{lahaye07}, as well as the recently realized 
polar molecules \cite{ni08}. An alternative class of strong interactions appears
between atoms excited into a Rydberg state \cite{singer05}.
%
%
In this letter, we show that the shape and character of the interaction potential between ground state atoms can be altered by
 dressing them with a Rydberg level giving rise to a collective many-body interaction.


The optical dressing of ground state atoms with an excited electronic state has
extensively been studied in the past in the context of  ``blue shield'' for the
supression of inelastic collisions \cite{weiner99}.
Motivated by recent experimental progress in manipulating Rydberg atoms also in
the ultra-cold regime \cite{heidemann08}, the possibility of controlling the
interaction via Rydberg dressed ground state atoms has been proposed for the
dipole-dipole interaction in the strongly interacting regime \cite{pupillo10},
and for inducing short range interactions in a Bose-Einstein condensate \cite{henkel10}.  
In contrast to the earlier proposals \cite{santos00}, these
approaches use a finite detuning in order to reduce losses
via spontaneous emission from the excited Rydberg state.



Here, we study the influence of Rydberg dressed interactions on the ground
state wave function  of a Bose-Einstein condensate. The combination of strong 
interactions between the Rydberg state and the detuning gives rise to a large 
effective range of the interaction potential, which exceeds the natural interparticle 
distance in cold atomic gases. Within this regime, we show that a crossover 
from a two-particle interaction potential to a collective many-body interaction takes place.
This collective many-body interaction appears as a consequence of the dipole/van der Waals blockade 
extensively studied in the past \cite{singer04,tong04,vogt06,heidemann07,raitzsch08}.
We provide a description for this collective interaction in a Bose-Einstein condensate and discuss 
its experimental signatures.



\begin{figure}[ht]
 \includegraphics[width= 1\columnwidth]{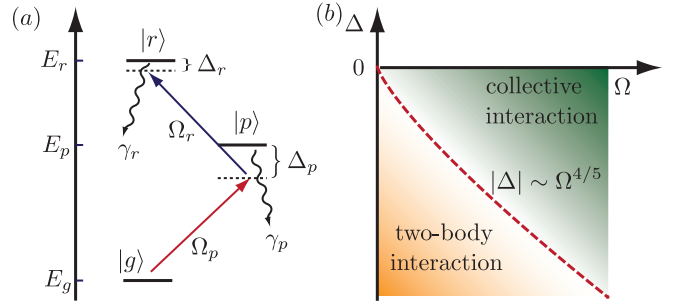}
 \caption{(a) Setup for the ground state dressing with the Rydberg level $|r\rangle$ via the intermediate $|p\rangle$ 
  state with the corresponding Rabi frequencies and detunings. (b) Diagram showing the crossover from the two-body 
  interaction potential into the regime with a collective many-body interaction. The crossover line is determined 
  by $\Delta \sim -\Omega^{4/5}$ (red dashed line), which preempts the condition of weak dressing $\Omega \lesssim |\Delta|$.}   \label{fig1}
\end{figure}


In the following, we are interested in a two-photon coupling of the ground
state atoms to a high lying Rydberg state; see \cite{heidemann08} for such an
experimental setup.  Then, the relevant internal structure for each atom is
given by a three-level system: the atomic ground state $|g\rangle_{i}$ is
coupled to the Rydberg level $|r\rangle_{i}$ via  the intermediate
$|p\rangle_{i}$ state, see Fig.~\ref{fig1}.  The Rabi frequency  and detuning
for the transition from the ground state to the p-level is denoted as
$\Omega_{p}$ and $\Delta_{p}$, and for the transition from the $p$-level to the
Rydberg state $\Omega_{r}$ and $\Delta_{r}$, respectively.  The intermediate $p$-level is far
detuned, $\Delta_{p}\gg \Omega_{r},\Omega_{p}$, and can be adiabatically
eliminated. Then the dynamics for a single atom reduces to an effective two
level system $H_{i} =  \hbar\left[ \Omega \sigma_{x}^{(i)}- \Delta \sigma_{z}^{(i)} \right]/2$
%
%
with the two-photon Rabi frequency $\Omega = \Omega_{r} \Omega_{p}/2
|\Delta_{p}|$ and the total detuning   $\Delta = \Delta_{r}+\Delta_{p}$; here $\sigma_{\alpha}^{(i)}$ denote the Pauli matrices.  
In the following, we first derive the two-body interaction
potential in the regime with large detuning, i.e., $\Omega  \ll | \Delta|$. Then,
the ground state atom is weakly dressed with the excited Rydberg level
$|+\rangle \approx(1- \Omega^2/8 \Delta^2) |g\rangle + \Omega/2 \Delta
|r\rangle$ providing a small fraction $f \approx  \Omega^2/4 \Delta^2$ of
excited Rydberg atoms. 
At large distances and in absence of F\"orster resonances \cite{vogt06}, the interaction 
between Rydberg states is dominated by the van der
Waals interaction $V_{\rs vdW}({\bf r}) = C_{6}/r^6$. In the following, we are
interested in a repulsive van der Waals interaction $C_{6}>0$ and red detuning
for the two-photon transition $\Delta<0$; the generalization to attractive
interactions with blue detuning is straightforward.  Then, the weak dressing of
the two-ground state atoms gives rise to the induced interaction $V_{\rs
eff}({\bf r}) \sim \Omega^4/ 16 \Delta^4 C_{6}/r^6$ at large distances $r \gg
\xi_0 \equiv \left(C_{6}/2 \hbar |\Delta|\right)^{1/6}$. 
On the other hand, for
short distances $r < \xi_0$, the two atoms enter the van-der Waals blockaded
regime, where only one atom is dressed with the Rydberg state, and the 
interaction saturates at an energy shift $\sim \Omega^4/\Delta^4 C_{6}/\xi_0^6$.
The full interaction potential $V_{\rs eff}$ is derived within the adiabatic Born Oppenheimer
approach  
$ V_{\rs eff}({\bf r}) =   \hbar \Omega^4/(8 | \Delta|^3) [1+ \left(r/\xi_0\right)^{6}]^{-1}$  \cite{pupillo10,henkel10},
which in Fourier space reduces to
\begin{equation}
 V_{\rs eff}({\bf q}) = \frac{\pi^2}{12} \frac{\hbar \Omega^4}{|\Delta|^3} \xi_0^3 f(\xi_0 q) \label{interactionpotential}
\end{equation}
with $f(z) = e^{-z/2}\left[ e^{-z/2} - 2\cos\left(\sqrt{3} z/2 + \pi/3\right) \right]/z$.
Within first Born approximation, this two-body interaction gives rise to  the
$s$-wave scattering length 
\begin{equation}
g_{\rs eff}= \frac{4 \pi \hbar^2  a_{\rs eff}}{m}= \frac{\pi^2}{12} \frac{\hbar \Omega^4}{| \Delta|^3} \xi_0^3. \label{geff}
\end{equation}
Note, that on short distances, additional crossings with the surrounding Rydberg levels appear, 
which give rise to very narrow resonances and opens up an additional loss chanel. However, due to 
the extremely narrow structure of these resonances one can expect that the particles mainly move 
diabatically across, and these losses are strongly suppressed; a detailed 
theoretical analysis of these losses is so-far missing.

In the many-body system, the effective interaction between the particles  can in general
also contain many-body interaction terms. For weak Rydberg dressing with $(\Omega/\Delta)^2 \ll 1$, 
these additional terms become relevant at high particle densities $n$:  first, the two-particle blockade radius $\xi_{0}$ gives rise
to a maximal density of excited Rydberg atoms $1/\xi_{0}^3$ above which collective phenomena become relevant. 
On the other hand, the density of excited Rydberg atoms due to the weak dressing is given by by $(\Omega/ 2 \Delta)^2 n$.
Consequently,  the validity of the two-body interaction $V_{\rs eff}({\bf r})$ is limited to the dilute 
regime
\begin{equation}
  n \xi_0^3 \ll \frac{4 \Delta^2}{\Omega^2} ,
\end{equation}
At higher densities $n \gtrsim n_{c} = 4 |\Delta|^{5/2}/(\Omega^2
\sqrt{C_{6}/2})$
a crossover into the collective regime takes place (see dashed line in
Fig.~\ref{fig1}) and the effective interaction is dominated by collective many-body interactions.  
The behavior $| \Delta|^5 \sim (C_{6} n^2\Omega^4)$ of this cross-over line derived within this
simple estimate agrees with the one expected from the universal scaling theory
\cite{weimer08,loew09}.

In the following, we provide a description of this effective interaction for a Bose-Einstein condensate.
The idea is to derive for a fixed atomic density $n$ the internal energy $E_{\rs eff}[n]$ of the driven Rydberg system within a Born-Oppenheimer approximation.  
This internal energy  then describes the collective interaction potential giving rise to a generalized Gross-Pitaevskii equation 
\begin{equation}
  i   \hbar \partial_{t} \psi(t,{\bf r}) = \left\{H_{0}  + g_{s} n(t,{\bf r})+E'_{\rs eff}\left[n(t,{\bf r})\right] \right\} \psi(t,{\bf r}) \label{GPequation}
\end{equation}
with the condensate wave function $\psi(t,{\bf r})$, the density $n(t, {\bf r})=|\psi(t,{\bf r})|^2$, and the derivative $E_{\rs eff}'[n]= \partial_{n}E_{\rs eff}[n]$.   Here, 
$H_{0}= - \hbar^2 \Delta/ (2 m)+V_{\rs ext}({\bf r}) - \mu$ describes the non-interacting part
with $V_{\rs ext}({\bf r})$ the external trapping potential, $m$ the mass of
the particles,  and $\mu$ the chemical potential, while $g_{s}$ accounts for the $s$-wave scattering between ground state atoms. This approach remains valid for momenta of the wave function 
smaller than the characteristic length scale for the effective interaction.

\begin{figure}[t]
 \includegraphics[width= 1\columnwidth]{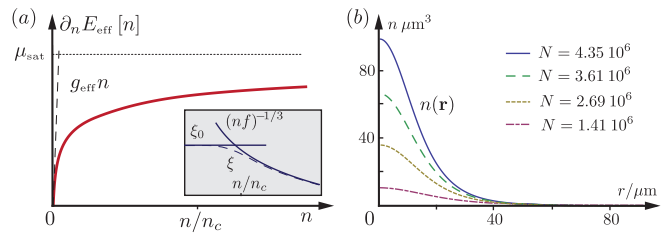}
 \caption{(a) Energy funcitonal $\partial_{n}E_{\rs eff}[n]$ for $|\Omega/\Delta| = 0.072$. The dashed line
 describes the behavior at low densities $E_{\rs eff}[n]= g_{\rs eff}n^2/2$ and its deviation indicates the early break down of the 
 two-particle interactions, while the dotted line accounts for the saturation at high densities, although this saturation is very slow.  
 The inset shows
 the behavior in the Blockade radius across the critical density $n_{c}$. (dotted line is guide to the eye). (b) Behavior of the density profile for a Bose-Einstein condensate 
 with $g_s = 0$ and the collective many-body interaction  $E'_{\rs eff}[n]$
 for increasing particle numbers in a harmonic trap with $\omega=20{\rm Hz}$. A characteristic feature is the strong increase of the central density $n(0)$ in contrast to the conventional Thomas-Fermi scaling $n(0) \sim N^{2/5}$.}   \label{fig2}
\end{figure}

The internal energy $E_{\rs eff}[n]$ can be derived within  a mean-field/variational
approach.  The Hamiltonian for the internal degree of freedom is described  by
a spin Hamiltonian  
\cite{heidemann07}
\begin{displaymath}
  H = \sum_i \left( - \frac{\hbar \Delta}{2} \sigma_z^{(i)} + \frac{\hbar \Omega}{2} \sigma_x^{(i)} 
  + \frac{C_6}{2}  \sum_{i\neq j}\frac{P^{(i)} P^{(j)}}{|{\bf r}_i -{\bf r}_j|^6} \right),
\end{displaymath}
with $P^{(i)} = (1+ \sigma_{z}^{(i)})/2$  the projection onto the excited
Rydberg state $|r\rangle$. 
The strong repulsion between the Rydberg atoms gives rise to the correlation
function $g({\bf r}_{i}-{\bf r}_{j}) = \langle  P^{(i)}
P^{(j)}\rangle/f^2$. The correlation function vanishes on short distances,
i.e., $g({\bf r}) \approx 0$ for $|{\bf r}| \ll \xi$,  due to the Blockade of
Rydberg excitations, and approaches unity $g({\bf r}) \rightarrow 1$ at large
distances $|{\bf r}| \gg \xi$. The characteristic correlation length scale $\xi$ denotes the 
collective blockade radius. Note, that this correlation function is independent on the
remaining particle positions due to the homogeneous distribution of the particles 
in the BEC. A variational wave function, which allows one to include the strong correlations
takes the form
\begin{equation}
   |0\rangle = \frac{1}{\mathcal{N}}\sum_{s_{1} \ldots s_{N}} \left[ \prod_{i \neq j}^{N} 
   C_{s_{i} s_{j}}({\bf r}_{i}- {\bf r}_{j}) \right]  | s_{1} \ldots s_{N}\rangle,
\end{equation}
where $C_{s s'}({\bf r})$ with $s \in \{g,r\}$
account for correlations on distances shorter than a characteristic length
scale $\xi$, and reduce to $C_{ss'} \rightarrow \alpha_{s} \alpha_{s'}$ for large distances with $\alpha_{g}= \cos\theta$ and $\alpha_{r} = \sin \theta$. 
Then, this
wave function describes a paramagnetic phase with the spins aligned along the
direction $\langle  \boldsymbol{\sigma}^{(i)}\rangle$, i.e., $\cos 2
\theta = -\langle \sigma^{(i)}_{z}\rangle $ and $\sin 2\theta = - \langle
\sigma^{(i)}_{x}\rangle $, and the fraction of excited Rydberg atoms reduces to
$f\equiv  \langle P^{(i)}\rangle = \sin^{2} \theta$. Note, that the relation between $C_{ss'}({\bf r})$
and the correlation function g({\bf r}) is highly non-trivial. In the following, we are interested in using the correlation length $\xi$
as a variational parameter, and therfore, the precise shape of $C_{ss'}$ is irrelevant.
The energy density $ \epsilon_{\rs var}(\theta, \xi)= \langle0| H|0\rangle$ for the internal degree of
freedom reduces to 
\begin{equation}
   \epsilon_{\rs var}(\theta, \xi) = \frac{\hbar \Delta n}{2} \cos 2 \theta  - 
    \frac{ \hbar \Omega n}{2} \sin 2 \theta  + \lambda \sin^4 \theta \frac{C_{6} n^2}{\xi^3}  \label{variationalenergy}.
\end{equation}
The variational parameters are the excited Rydberg fraction $f=\sin^2
\theta$ and the correlation length $\xi$. 
The dimensionless parameter $\lambda / \xi^3 = \int d^3x g(\mathbf{x})/|\mathbf{x}|^6 = 
1/\xi^3 \cdot  \int d^3x g(\xi \mathbf{ x})/|\mathbf{x}|^6$ is determined by the details of 
the correlation function.  It is important to note, that the qualitative behavior of the energy 
functional is independent on the precise value of $\lambda$, and therefore, this parameter will be fixed 
by the constraint to reproduce the correct asymptotical behavior at low densities. 
The dressing of the ground state atoms with the excited Rydberg level leads to an energy shift
$\epsilon_{0}= -\hbar n \sqrt{\Delta^2+\Omega^2}/2$ even in absence of Rydberg
interactions. Subtracting this  contribution, the energy functional $E_{\rs eff}[n]$, describing 
the Rydberg dressed interactions in the collective regime, is obtained by the energy for the state 
adiabatically connected to the non-interacting system: this wave function is obtained by minimizing
the energy function $\epsilon_{\rs var}$, i.e.,
\begin{equation}
E_{\rs eff}[n] = \min_{\{|0\rangle\}}\left[\epsilon_{\rs var}(\theta, \xi)\right] - \epsilon_{0} ,
\end{equation}
for fixed ground state density $n$, Rabi frequency $\Omega$, and detuning
$\Delta$. 
The value $\theta$ for the minimal energy density follows
from the equation $\partial_{\theta} \epsilon_{\rs var}(\theta, \xi) = 0$;
note, that this condition is equivalent to the self-consistency equation
derived in Ref.~\cite{weimer08}.  On the other hand, the last term in equation 
(\ref{variationalenergy}) always gives a positive contribution to the energy functional, 
and it will be minimized by maximizing the correlation length $\xi$.  However, the correlation length 
satisfies several constaints: in the weakly interacting regime the correlations due to the Rydberg excitations 
are limited by the two-particle Blockade radius $\xi_0$ \cite{heidemann07}. Entering the collective regime 
the distance between the Rydberg atoms is  further reduced and the correlations are bounded by  $(1/n f)^{1/3}$, 
i.e. $\xi \approx (1/n f)^{1/3}$, see inset in Fig.~\ref{fig2}. 
Note, that the qualitative behavior of the system is independent on
the precise choice of $\xi$ but only weakly renormalized the dimensionless parameter $\lambda$.

The energy functional $E_{\rs eff}[n]$  is shown in Fig. 2 for varying density n.
In the low density regime, $n \ll n_{c} = 4 |\Delta|^{5/2}/(\Omega^2
\sqrt{C_{6}/2})$,  it reduces to a pure two-particle interaction 
$E_{\rs eff} = g_{\rs eff} n^2/2$.  The comparison  with the exact two-body interaction Eq.~(\ref{geff}) 
allows us to fix the dimensionless parameter $\lambda$ and within first Born approximation it takes the form $\lambda = 2 \pi^2 /3$.
Strong deviation from the two-body interaction can already be seen
at densities as low as $0.05  n_c$, see Fig.~\ref{fig2}. Collective blockade phenomena give rise to
a very broad crossover, with the energy functional $E_{\rs eff}[n]$ eventually saturating
at a chemical potential $E_{\rs eff}[n] \approx n \mu_{\rs sat}=n \hbar \left(
\sqrt{\Delta^2 +\Omega^2} -|\Delta| \right)/2$ at high densities $n \gg n_{c}$.
This behavior can be understood from the following argument: (i) each atom within
a Blockade radius gives rise to an energy shift in the functional, and (ii) in the high density limit
to fraction of excited Rydberg atoms vanishes. Therefore, all atoms up to a non-extensive part are within
a Blockade radius, and thus basically free.

\begin{figure}[ht]
 \includegraphics[width= 0.8 \columnwidth]{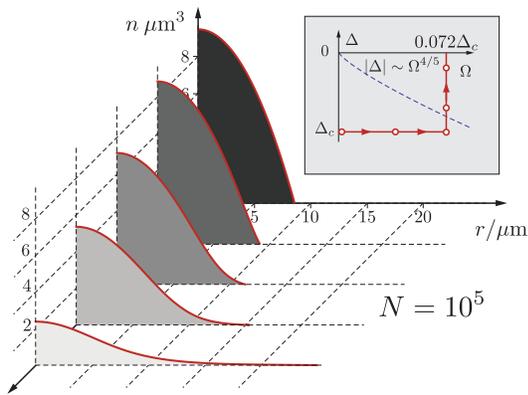}
 \caption{Density profile $n(r)$ for a Bose-Einstein condesate of $^{87}{\rm
Rb}$ atoms  in a harmonic trap $\omega=100{\rm Hz}$ with background scattering
length $a_{s}=5.7 {\rm nm}$ for increasing influence of the Rydberg dressed
interaction.  The inset shows the path within the parameter space for
increasing interaction strengths.}   \label{fig3}
\end{figure}

Finally, we can derive the ground state wave function for a Bose-Einstein
condensate in an harmonic trap with the influence of the Rydberg dressed
interaction within the Thomas-Fermi approximation.  
The density profile for  the collective many-body interaction
is shown in Fig.~\ref{fig2}(b) for increasing particle number $N$.
A characteristic feature is the strong increase of the density in the trap center. 
This behavior is a consequence of the linear behavior of the collective interaction
 potential at high particle densities $n \gg n_c$, and allows one to distinguish the 
 collective interaction from 
a two-body interaction or heating effects.
The density profile for an
atomic condensate of $^{87}{\rm Rb}$ atoms with a background scattering length
$a_{s}=5.7 {\rm nm}$ and fixed particle number $N=10^5$ is shown in
Fig.\ref{fig3}.  For weak Rydberg dressed interactions, the density profile is
given by the inverted parabola profile $n({\bf r})=\left[ \mu - V_{\rs ext}({\bf r}) \right]/g_\text{eff}$.  
The crossover into the collective interaction regime is again signaled by the characteristic change 
in the shape of the atomic cloud.


In the following, we will determine the parameters for an experimentally realistic setup.
For weak dressing, the population of the Rydberg  state 
is suppressed by the
factor $(\Omega/\Delta)^2 $, as well as the intermediate $p$-level 
by the
factor $(\Omega_{p}/\Delta_{p})^2$. Denoting the decay rate of the Rydberg state and the $p$-level with $\gamma_r$ and $\gamma_p$, respectively, the total decay rate $\gamma_{\rs
eff}$ for spontaneous emission is
given by
\begin{equation}
\gamma_{\rs eff} 	
=  \frac{\Omega_{p}^{2}}{4\Delta_{p}^{2}} \left( \gamma_{p}+ \frac{\Omega_{r}^2}{4 \Delta^2} \gamma_{r}\right).
\end{equation}
The detuning $\Delta_{p}$ and Rabi frequency $\Omega_{p}$
of the laser coupling the ground state and the intermediate $p$-level only
determine the global life time via its ratio $(\Omega_{r}/\Delta_{r})^2$.
Second, the Rabi frequency $\Omega_{r}$ for the coupling to the Rydberg state is
limited by experimental restrictions on available laser power.  Consequently, the strongest influence 
of the Rydberg dressed interaction is obtained by the choice of the detuning $\Delta$, which extremizes 
the induced $s$-wave scattering length for fixed life time of the system, i.e.,  $\partial_{\Delta} g_{\rs eff} = 0$.
Then, the optimal detuning satisfies  $ \Delta_{c} = -|\Omega_{r}|\sqrt{\gamma_{r}/(28  \gamma_{p})} $ with the induced
$s$-wave scattering length
\begin{displaymath}
  a_{\rs eff} =\frac{m}{4 \pi \hbar^2} g_{\rs eff}(\Delta_{c})=\frac{m}{4 \pi \hbar^2}  \frac{49 \pi^2}{48}
  \sqrt{C_{6}  /2} \frac{\gamma^2}{\gamma_{r}^2} \left(\frac{\Omega_{r}^2\gamma_{r}}{28 \gamma_{p}} \right)^{1/4} .
\end{displaymath}
For ${{}^{87}\rm
Rb}$ weakly dressed with the Rydberg level $|35s\rangle$,
the van der Waals interaction takes the strength
$C_{6} = 1.31 \:  10^{18} {\rm a.u.}$ \cite{singer04} ,
and the decay rates $\gamma_{r} = 4 {\rm kHz}$ and $\gamma_{p}=6{\rm
MHz}$. For a realistic experimental setup, Rabi frequencies with $\Omega_{r}=22
{\rm MHz}$ have been reached. Finally, a life time of $\gamma =6
{\rm Hz}$ is sufficient for a fast experiment, and we obtain
\begin{equation}
   a_{\rs eff} = 1.36 \:  \gamma^2 \: {\rm nm} {\rm Hz}^{-2} = 49.5 {\rm nm},
\end{equation}
compared to $a_s = 5.7 {\rm nm}$, which is an increase of the scattering length by an order of magnitude.
The optimal detuning reduces to $|\Delta_{c} |= 107 {\rm kHz}$ with the
characteristic length scale $\xi_0 = 3 \mu {\rm m}$. The coupling to the
$p$-level requires the strength $|\Omega_{p}/\Delta_{p}|= \sqrt{\gamma/2
\gamma_{p}}=7.0 \: 10^{-4}$, and the effective Rabi frequency becomes $\Omega =
7.8 {\rm kHz} $, thus $|\Omega/\Delta_{c}|=\sqrt{7 \gamma/ 2 \gamma_{r}} = 0.072$.
The critical density for the crossover from the two-body interaction to the collective many-body interaction reduces  to
$n_{c}= 2.57 \: \:10^{13} {\rm cm}^{-3}$, and consequently, for typical peak densities 
$n_\text{peak} \sim 10^{14} {\rm cm}^{-3}$ of a typical BEC experiment the system is well within the collective regime, and allows
for the detection of this novel interaction.
Note, that within  the collective regime, the number of excited Rydberg
atoms is further reduced due to the Blockade phenomena. This in principle allows one to further reduce the 
detuning without increasing the losses from spontaneous emission.
However, in a trapped system, this will increase the losses at the edge of the condensate, 
where the density drops below the critical density $n_c$.




Finally, we would like to point out, that the weakly dressed Rydberg interaction could be further tuned by a micro-wave field coupling different Rydberg levels \cite{Muller08}.
%
%
%
As opposed to the much stronger van der Waals repulsion between the Rydberg levels, the micro-wave field only 
gives rise to a weak dipole-dipole interaction, which becomes relevant on large distances $r \gg \xi$ due to its slow 
decay. As a consequence,  it does not change the behavior on distances comparable to the Blockade radius. 
Such a scenario is in analogy  to the possibility to tune interaction potentials with cold polar molecules  \cite{Gorshkov08},
and will allow one to explore the interesting properties of anisotropic dipole-dipole interactions in the collective regime.


%



Support from the DFG (Deutsche Forschungsgemeinschaft) within
SFB/TRR21 is acknowledged.


\end{document}